\documentclass{article}

\usepackage{PRIMEarxiv}

\usepackage[utf8]{inputenc} 
\usepackage[T1]{fontenc}    
\usepackage{hyperref}       
\usepackage{url}            
\usepackage{booktabs}       
\usepackage{amsfonts}       
\usepackage{nicefrac}       
\usepackage{microtype}      
\usepackage{lipsum}
\usepackage{fancyhdr}       
\usepackage{graphicx}       
\graphicspath{{media/}}     

\usepackage{listings}
\usepackage{amsthm}
\usepackage{amssymb}
\usepackage{subfigure}
\usepackage{multicol}
\usepackage{ctable}
\usepackage{float}
\usepackage{times}
\usepackage{pifont}
\usepackage{upgreek}
\usepackage{algorithmicx}
\usepackage{algorithm}
\usepackage{algpseudocode}
\usepackage{multirow}
\usepackage{caption}

\title{End-to-End QoS for the Open Source Safety-Relevant RISC-V SELENE Platform
}

\author{
  Pablo Andreu$^\S$, Carles Hernandez$^\S$, Tomas Picornell$^\S$, Pedro Lopez$^\S$, Sergi Alcaide$^{\dagger,\ddagger}$, Francisco Bas$^{\dagger,\ddagger}$, \\
  \textbf{Pedro Benedicte$^\dagger$, Guillem Cabo$^\dagger$, Feng Chang$^\dagger$, Francisco Fuentes$^\dagger$, Jaume Abella$^\dagger$}\\
  $^\S$Universitat Politecnica de Valencia (UPV), Spain\\
  $^\dagger$Barcelona Supercomputing Center (BSC), Spain\\
  $^\ddagger$Universitat Politecnica de Catalunya (UPC), Spain\\
}

\begin{document}
\maketitle

\begin{abstract}
This paper presents the end-to-end QoS approach to provide performance guarantees followed in the SELENE platform, a high-performance RISC-V based heterogeneous SoC for safety-related real-time systems. Our QoS approach includes smart interconnect solutions for buses and NoCs, along with multicore interference-aware statistics units to, cooperatively, achieve end-to-end QoS.
\end{abstract}


\section{Introduction}

Increasing performance demands for safety-related systems impose the adoption of higher-performance multi-processor systems-on-chip (MPSoCs). However, those MPSoCs include medium or large multicores with cache memories, accelerators and memory controllers, which need being shared dynamically and at fine-grain (e.g. below microsecond scale) for efficiency reasons. This clashes against common practice where few hardware resources are shared, and sharing occurs at a coarse granularity (e.g. above millisecond scale) so that their sharing can be managed at software level by the hypervisor or the real-time operating system (RTOS). 

To address this emerging concern, hardware must provide appropriate support to guarantee performance controllability so that resources are shared efficiently and \textbf{\emph{fairly}}. In particular, performance concerns such as starvation, deadline overruns for real-time tasks, and priority inversion must be avoided by construction, and this can only occur if the MPSoC provides enough Quality-of-Service (QoS) knobs to guarantee a reasonable use of resources to all cores, in line with user-level requirements.

Some solutions such as resource partitioning (e.g. cache partitioning), and fair and programmable arbitration policies (e.g. round robin, priority-based) have been explored in the literature to address specific QoS concerns in platforms considered for safety-related applications. This has been studied, for instance, in the case of the Xilinx Zynq UltraScale+~\cite{QoSECRTS}, a platform of interest for avionics industrials among others~\cite{ZUSRC}. Unfortunately, in general, those solutions do not provide easy ways to manage end-to-end QoS so that uncontrolled sharing in a given hardware resource defeats the informed sharing performed in others.

This paper presents an end-to-end QoS approach to provide performance guarantees, integrated in the SELENE platform, a high-performance RISC-V based MPSoC for safety-related real-time systems~\cite{SELENEgit}. Our QoS approach includes smart interconnect solutions for buses and NoCs, along with multicore interference-aware statistics units to, cooperatively, achieve end-to-end QoS, hence enabling software layers with means to guarantee that deadlines are met for real-time tasks, critical tasks do not experience starvation, and priorities and performance requirements can be met without priority inversion for mixed-criticality systems.

\section{Baseline SoC}

\begin{figure}[ht!]
    \centering
    \includegraphics[width=0.7\textwidth]{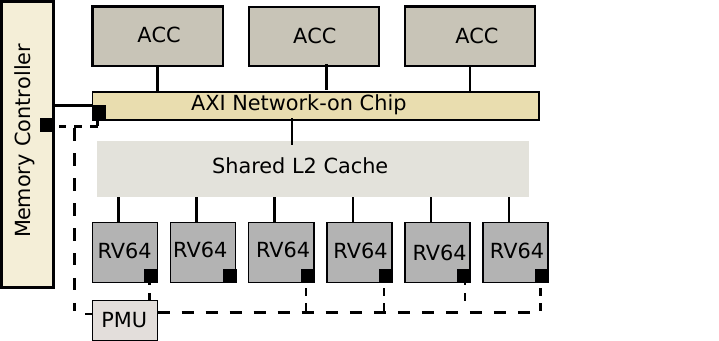}
    \caption{SELENE Baseline SoC}
    \label{fig:SoC}
\end{figure}

The SELENE System-on-Chip (SoC) comprises NOEL-V RISC-V cores and an AI accelerator subsystem. The SELENE SoC has six 64-bit NOEL-V cores with private L1 data and instruction caches that access to a shared L2 cache using an AMBA AHB on-chip bus. Cores and accelerators are connected to an AXI network-on-chip (NoC) to access memory and other SoC peripherals. This hierarchical interconnect allows maintaining coherence across cores at the bus level using a simple snoopy-based protocol. Figure~\ref{fig:SoC} shows a simplified view of the SELENE SoC. 

From a QoS perspective, the on-chip bus, the AXI NoC and the memory controller are the most critical components of the SELENE SoC. In the following sections, we detail the modifications required in these components to implement our end-to-end QoS approach. 

\section{End-to-End QoS Approach}

This section introduces our approach to achieve end-to-end QoS management. In particular, we introduce the overall approach, and then we detail individual components.

\subsection{General Approach}

Our end-to-end QoS management builds on some key strategies:
\begin{itemize}
    \item Controlled resource utilization to achieve some form of time partitioning for interconnects and computing components.
    \item Classical space partitioning for storage shared components.
\end{itemize}

For the latter, space partitioning, we resort to usual solutions such as cache space partitioning for shared caches, as this has been proven to be convenient in several high-performance MPSoCs for critical real-time applications~\cite{SurveyMulticores}.

For the former, time partitioning, we rely on two key observations: (1) utilization of shared resources, such as for instance, interconnects, is not necessarily proportional to transactions~\cite{MladenDATE17}. Instead, different transactions may occupy shared resources for highly different time intervals, and hence, lead to different overall utilization levels despite having a similar number of transactions across contenders. (2) It is key determining the task (i.e. the core) responsible for each activity in the MPSoC that may cause contention on others. However, induced activities that are not exposed directly by the cores (e.g. a transaction between the L2 cache and DRAM) may lack owner identification, which challenges blaming the core causing potential interference on others if owner IDs are not conveniently propagated.

To address time partitioning in a holistic manner across the SoC, we rely on several key strategies and components as follows:
\begin{enumerate}
    \item Tracking activity ownership all along the MPSoC. In particular, we leverage a core ID mechanism analogous to Worldguard~\cite{WorldGuard}, initially intended for security purposes, to propagate core IDs all along the MPSoC so that the owner of each request is known anywhere, and is easy to tell who is creating contention on whom.
    \item Measuring contention caused and experienced everywhere. We leverage the SafeSU~\cite{SafeSU}, a multicore interference-aware statistics unit, to track how much contention each core causes on each other all along the MPSoC building on the core IDs provided by the Worldguard-like solution.
    \item Implementing programmable quota allocation mechanisms and contention-aware arbitration schemes in the interconnects to guarantee that time partitioning adheres to end user requirements so that time budgets are respected. At the same time, mechanisms to avoid starvation for offending cores (e.g. those trying to exceed their assigned quotas) are also deployed.
\end{enumerate}

Next, we describe the key technologies leveraged for the main shared components needing time partitioning.

\subsection{Bus-level Solutions}

\begin{figure}[ht!]
\centering \includegraphics[width=0.60\textwidth]{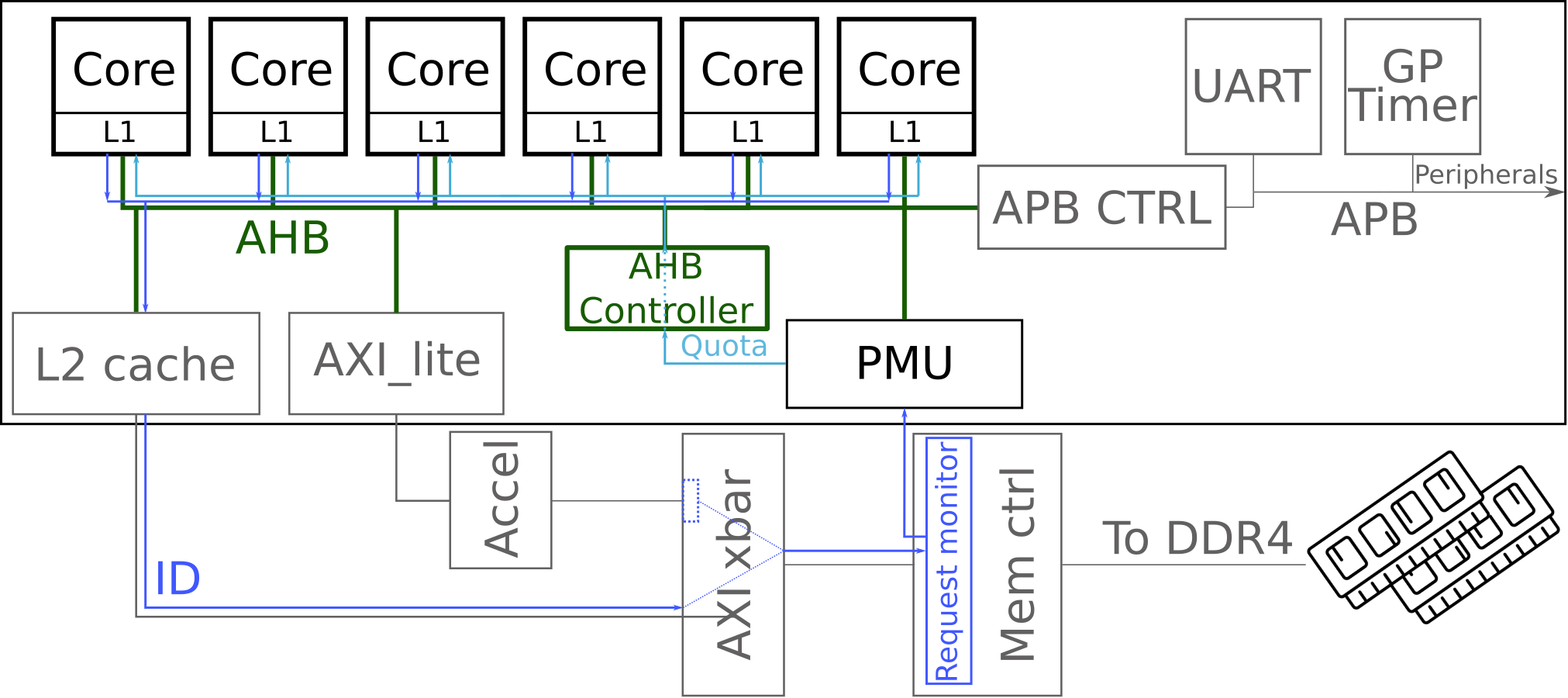}
\caption{SELENE SoC Hardware Support for QoS }
\label{fig:QoS}
\end{figure}

As explained before, cores are interconnected with an on-chip bus to the shared L2 cache. As illustrated in Figure~\ref{fig:QoS}, the on-chip bus in our baseline SoC implements a round-robin policy. Contention in the on-chip bus is measured by the SafeSU~\cite{SafeSU} module, also referred to as Performance Monitoring Unit (PMU), which snoops bus activity to accurately track the contention suffered (and caused) by each core. When the SafeSU detects that a core has exhausted a predefined contention quota, we have two alternatives to resolve the situation. The first alternative is to trigger an interrupt that has to be captured by a monitoring task. The other alternative that we refer to as hardware quota is to drive this signal to the on-chip bus arbiter to stall the activity of the offending core with hardware-only means. The actual enforcement mechanism depends on the needs of the application that is deployed in the system. 

\subsection{NoC-level Solutions}

The NoC interconnects cores, accelerators and the rest of peripherals with memory. Currently, the interconnect uses a crossbar-based structure where the masters, mostly cores and accelerators, contend to access the shared resources (memory devices). As in the case of the on-chip bus, the default arbitration policy is round-robin. 

To monitor contention and/or to implement arbitration decisions in a smarter way, we need to know who is responsible for each NoC request. We use the QoS bits of the AXI interface to retrieve this information in the NoC and subsequent SoC modules. Accelerators can directly include the initiator ID in the AXI requests. Requests from cores to the NoC share the same AXI link. Thus, we cannot directly infer who is the initiator of the request. To solve this, the L2 cache module, or an AHB to AXI bridge in the absence of an L2 cache, injects the initiator ID in the AXI QoS bits before forwarding the request to the NoC. 

\subsection{Memory Controller-level Solutions}

A memory transaction may cause interference to one or multiple requests. Therefore, we need to identify requests interference at memory controller to blame the corresponding owner. To do so, we rely on the AXI request information that we propagate in every request. 

\begin{figure}[ht!]
\centering \includegraphics[width=0.60\textwidth]{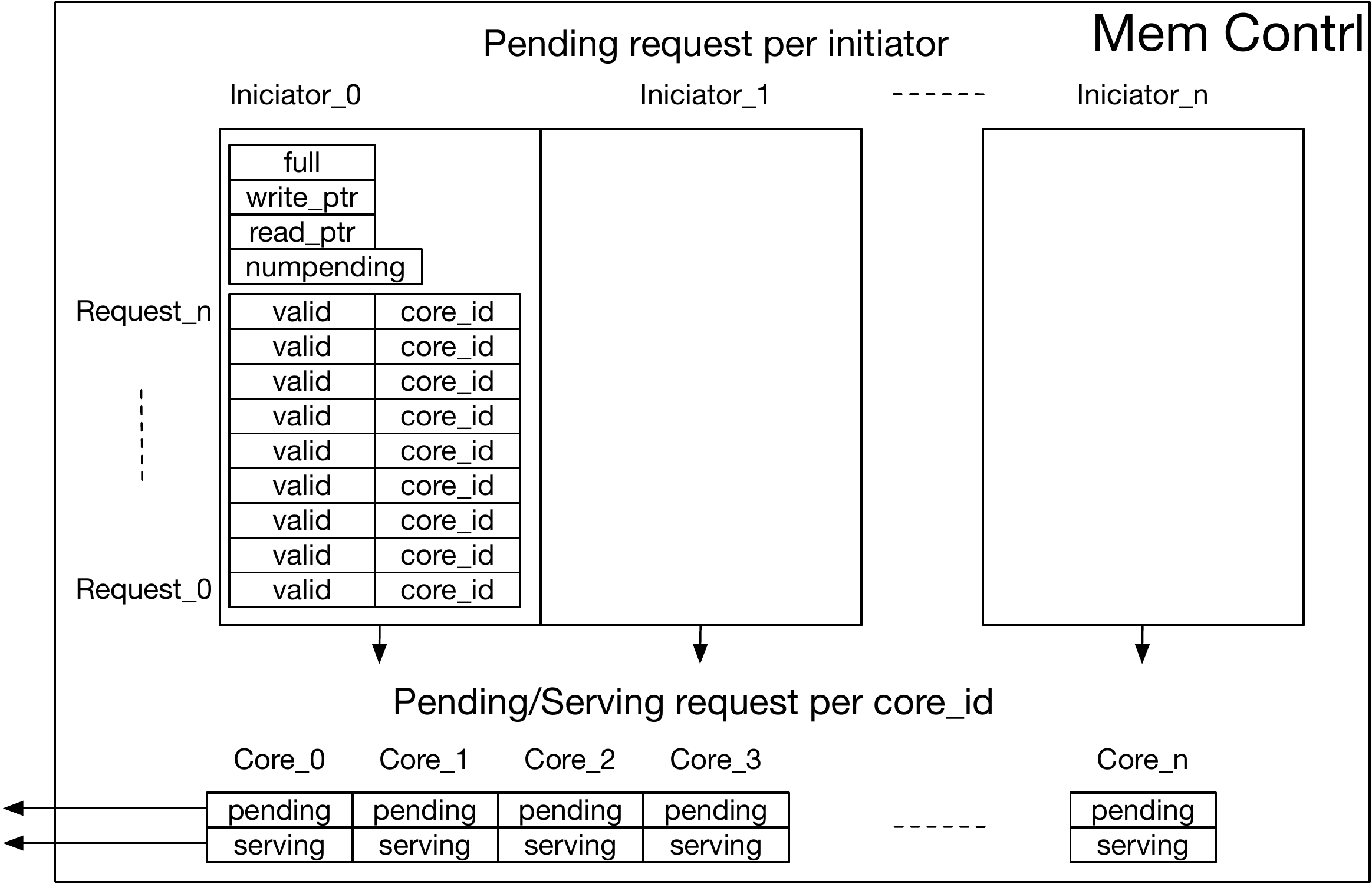}
\caption{{Memory controller signals to monitor pending and serving memory transactions. These structures are implemented for read and write request separately.}}
\label{fig:Memory_controller_signals}
\end{figure}

At the memory controller, memory requests arrive and are identified by the initiator ID. In order to quantify contention, we keep track of current pending and serving memory transactions. To do so, we extend the architecture of the memory controller to define some information structures as shown in Figure~\ref{fig:Memory_controller_signals}.

In the top-left part of Figure~\ref{fig:Memory_controller_signals}, we show the FIFO queues to store pending requests for every initiator. These queues keeps track of pending request order and allow us to easily identify the involved owner ID. The lower part of Figure~\ref{fig:Memory_controller_signals} illustrates the status of the memory request for every core. This information is propagated to the SafeSU~\cite{SafeSU} to quantify the contention each core is causing.

\section{Summary}

This paper describes the SELENE end-to-end approach to achieve QoS in heterogeneous multicores and reviews the current status of the implementation of this approach in the SELENE SoC.

\section*{Acknowledgments}
This work has received funding from the European Union's Horizon 2020 research and innovation programme under grant agreement no. 871467. 
BSC work has also been partially supported by the Spanish Ministry of Science and Innovation under grant PID2019-107255GB-C21/AEI/10.13039/501100011033.

\bibliographystyle{plain}  
\bibliography{biblio.bib}

\begin{thebibliography}{1}

\bibitem{SafeSU}
Guillem Cabo, Francisco Bas, Ruben Lorenzo, David Trilla, Sergi Alcaide, Miquel
  Moretó, Carles Hernández, and Jaume Abella.
\newblock Safesu: an extended statistics unit for multicore timing
  interference.
\newblock In {\em 2021 IEEE European Test Symposium (ETS)}, pages 1--4, 2021.

\bibitem{SELENEgit}
{H2020 SELENE consortium}.
\newblock {SELENE RISC-V open source hardware platform}.
\newblock \url{https://gitlab.com/selene-riscv-platform}, 2021.

\bibitem{SurveyMulticores}
J.~{Perez-Cerrolaza et al.}
\newblock Multi-core devices for safety-critical systems: A survey.
\newblock {\em ACM Comput. Surv.}, 53(4), August 2020.

\bibitem{QoSECRTS}
Alejandro Serrano-Cases, Juan~M. Reina, Jaume Abella, Enrico Mezzetti, and
  Francisco~J. Cazorla.
\newblock {Leveraging Hardware QoS to Control Contention in the Xilinx Zynq
  UltraScale+ MPSoC}.
\newblock In Bj\"{o}rn~B. Brandenburg, editor, {\em 33rd Euromicro Conference
  on Real-Time Systems (ECRTS 2021)}, volume 196 of {\em Leibniz International
  Proceedings in Informatics (LIPIcs)}, pages 3:1--3:26, Dagstuhl, Germany,
  2021. Schloss Dagstuhl -- Leibniz-Zentrum f{\"u}r Informatik.

\bibitem{WorldGuard}
{SiFive}.
\newblock {SiFive Shield: An Open, Scalable Platform Architecture for
  Security}.
\newblock
  \url{https://www.sifive.com/blog/sifive-shield-an-open-scalable-platform-architecture},
  2019.

\bibitem{MladenDATE17}
Mladen Slijepcevic, Carles Hernandez, Jaume Abella, and Francisco~J. Cazorla.
\newblock Design and implementation of a fair credit-based bandwidth sharing
  scheme for buses.
\newblock In {\em Design, Automation Test in Europe Conference Exhibition
  (DATE), 2017}, pages 926--929, 2017.

\bibitem{ZUSRC}
XILINX.
\newblock {Rockwell Collins Uses Zynq UltraScale+ RFSoC Devices in
  Revolutionizing How Arrays are Produced and Fielded: Powered by Xilinx}.
\newblock
  \url{https://www.xilinx.com/video/corporate/rockwell-collins-rfsoc-revolutionizing-how-arrays-are-produced.html},
  2018.

\end{thebibliography}

\end{document}